\documentclass[trackchanges]{aastex701}
\usepackage{subcaption}
\usepackage{graphicx}  
\usepackage{natbib}
\usepackage{fancyhdr} % 用于自定义页脚
\usepackage{amssymb}
\usepackage{xcolor}
\usepackage{array}
\usepackage{caption}
\defcitealias{2026arXiv260616355O}{Paper I}

\newcommand\amm{NH$_{3}$}

\newcommand\kms{km s$^{-1}$}
\newcommand\water{H$_{2}$O}
\newcommand\jyb{Jy beam$^{-1}$}

\begin{document}
\shorttitle{Water maser}
\shortauthors{Chao et al.}
\title{Water maser detections toward 70 $\mu$m infrared dark clumps at early star formation stage}
\author[orcid=0000-0003-0128-4570,sname='China']{Chao Ou}
% \altaffiliation{Guangxi Key Laboratory for Relativistic Astrophysics}
\affiliation{Guangxi Key Laboratory for Relativistic Astrophysics, School of Physical Science and Technology, Guangxi University, Nanning 530004, PR China}
\email[no]{chaoou@st.gxu.edu.cn}  

\author[orcid=0000-0003-1275-5251,sname='China']{Shanghuo Li}
\affiliation{School of Astronomy and Space Science, Nanjing University, 163 Xianlin Avenue, Nanjing 210023, People’s Republic of China.}
\affiliation{Key Laboratory of Modern Astronomy and Astrophysics (Nanjing University), Ministry of Education, Nanjing 210023, People’s Republic of China.}
% \email[]{shanghuo.li@gmail.com}
\email{shli@nju.edu.cn}

\author[orcid=0000-0001-6106-1171, sname='China']{Junzhi Wang} 
% \altaffiliation{Guangxi Key Laboratory for Relativistic Astrophysics}
\affiliation{Guangxi Key Laboratory for Relativistic Astrophysics, School of Physical Science and Technology, Guangxi University, Nanning 530004, PR China}
\email{junzhiwang@gxu.edu.cn}

\author[orcid=0000-0003-3017-4418]{Ian W. Stephens} 
\altaffiliation{}
\affiliation{Department of Earth, Environment, and Physics, Worcester State University, Worcester, MA 01602, USA}
\affiliation{Center for Astrophysics—Harvard \& Smithsonian, 60 Garden Street, Cambridge, MA 02138, USA}
\email{}

\author[orcid=0000-0001-6106-1171, sname='China']{Yuqiang Li} 
\altaffiliation{}
\affiliation{Korea Astronomy and Space Science Institute, No. 776, Daedeok-daero, Yuseong-gu, Daejeon, Republic of Korea}
\email{}

% \author{et al.,}
% \affiliation{}
% \email{}

\correspondingauthor{Shanghuo Li and Junzhi Wang}
\email{shli@nju.edu.cn, junzhiwang@gxu.edu.cn }

% \begin{document}
% \maketitle

% \clearpage
% \end{document}
% \author[0000-0000-0000-0003,sname=Asia,gname=Mountain]{Asia Mountain}
% \altaffiliation{Astrosat Post-Doctoral Fellow}
% \affiliation{Tata Institute of Fundamental Research, Department of Astronomy}
% \email{fakeemail5@google.com}

% \author[0000-0000-0000-0004]{Coral Australia}
% \affiliation{James Cook University, Department of Physics}
% \email{fakeemail6@google.com}

% \author[gname=IceSheet]{Penguin Antarctica}
% \affiliation{Amundsen–Scott South Pole Station}
% \email{fakeemail7@google.com}

% \collaboration{all}{The Terra Mater collaboration}

\begin{abstract}

Water masers provided powerful probes of shocked gas and protostellar activity during the early stages of star formation. Using VLA K-band observations of the 22~GHz water line, we \textcolor{black}{report} the first detections of water masers toward dense cores
in \textcolor{black}{five} 70~$\mu$m infrared dark clumps, providing evidence for active star formation at an early evolutionary stage. We detected ten unresolved maser spots comprising thirteen velocity components, with peak flux densities of 13--499~mJy and deconvolved full widths at half maximum (FWHMs) of 0.51--3.54~\kms. Eleven components had FWHMs below 1~\kms, while their velocity offsets from the systemic velocities traced by the \amm(1,1) main line ranged from 0.83 to 50.73~\kms. \textcolor{black}{Four water maser spots} were associated with 1.3~mm continuum emission, three of which were also associated with molecular outflows traced by CO~2 -1 emission. No CO outflow signature was detected toward the remaining spots within the observed field of view, suggesting that water masers could trace deeply embedded protostellar activity whose outflows were not yet detectable in CO emission. For AGAL031, the non-detection of Stokes~V emission toward the strongest maser component yields an upper limit on the line of sight magnetic field strength on the order of \textcolor{black}{a few hundred mG}. Our results showed that star formation was already active in 70~$\mu$m infrared dark clumps and that 22~GHz water masers provided insights into early protostellar environments that remained difficult to probe with common star formation tracers.

\end{abstract}



\section{Introduction}
\label{sec:intro}

Protostellar outflows represent one of feedback mechanisms in the self-regulation of star formation \citep{2007ARA&A..45..565M, 2016ARA&A..54..491B, 2020SSRv..216...50C}. These outflows and jets are ubiquitous in both low- and high-mass star-forming regions (SFRs), playing a critical role in shaping the surrounding environment throughout the evolution of molecular clouds \citep{1992A&A...255..293F, 2010ApJ...721..222B, 2019A&A...631A..74M, 2019ApJ...886..130L, 2019ApJ...878...29L, 2019ApJ...886...36S,2020ApJ...903..119L}. Water masers serve as powerful alternative tracers, probing protostellar outflows, protoplanetary disks, and jets that significantly influence the surrounding molecular gas during early protostellar evolution \citep{1981ARA&A..19..231R, 1989ApJ...346..983E, 2002A&A...390..289B, 2014A&A...564L..11L, 2019A&A...631A..74M, 2021AJ....162...68O, 2025A&A...701A.167H}. While water is primarily sequestered as ice on dust grains in cold regions of SFRs, it can be released into gas phase and collisionally pumped by neutrals within shocked material, resulting in maser emission \citep{1989ApJ...346..983E, 2012ApJ...759L..37C, 2014MNRAS.440.1844S}. The most frequently observed transition is the $6_{16}-5_{23}$ rotational line at $22\text{ GHz}$ in K-band, first discovered in the interstellar medium by \citet{1969Natur.221..626C}. The $22\text{ GHz}$ water maser effectively traces shocks driven by the interaction between protostellar outflows and circumstellar material, making it an essential tool for studying deeply embedded objects that are actively dispersing their surrounding molecular gas \citep{2006ApJ...651L.125W,2011A&A...527A..41S,2014MNRAS.439.3275W}. Thus, water masers are highly complementary to tracers like CO in probing the protostellar environment.

 Strong outflows or feedback from SFRs produce shocks that impact the dense gas surrounding the protostar \citep{2005Ap&SS.295...71I, 2016ARA&A..54..491B, 2019A&A...631A..74M}. As direct probes of these shocked environments, water masers can be utilized to constrain local magnetic field strengths \citep{2002ApJ...580..928S}. Beyond tracing kinematics, the Zeeman splitting of water masers provides a direct measurement of magnetic field strengths in SFRs \citep{1989A&A...214..333F, 2019FrASS...6...66C}. 
 %For instance, Karl G. Jansky Very Large Array (VLA) polarimetric observations of $22\text{ GHz}$ water masers have measured line-of-sight magnetic field strengths ($B_{\rm LOS}$) of $\approx 113 $ milligauss($\text{mG}$) in the low-mass system IRAS 16293-2422, and between $123$ and $156\text{ mG}$ in the high-mass object IRAS 19035+0641 A \citep{2012A&A...542A..14A, 2024ApJ...962..133R}. 
 \textcolor{black}{For instance, polarimetric observations of $22\text{ GHz}$ water masers with the Karl G. Jansky Very Large Array (VLA) measured a line-of-sight magnetic field strength ($B_{\rm LOS}$) of $\sim 113 $ milligauss($\text{mG}$) toward the low-mass system IRAS 16293$-$2422 \citep{2012A&A...542A..14A}, and values of $123\text{ mG}$ and $156\text{ mG}$ toward the high-mass object IRAS 19035+0641 A \citep{2024ApJ...962..133R}.} Nevertheless, magnetic field strengths in shock-driven environments during the earliest stages of star formation remain poorly understood, primarily due to the low detection rate of suitable water masers \citep{2022AJ....163..124L}.

In this work, we present VLA K-band observations of five $70\text{ }\mu\text{m}$ infrared dark clumps, targeting the $22\text{ GHz}$ water maser and ammonia transitions. The observations are described in Sect. \ref{sec:obs}. In Sect. \ref{sec:result}, we present the observational results of water masers.  We study the spatial association between water maser and CO outflows previously identified in \cite{2019ApJ...886..130L}, and utilize the water maser peak intensity to constrain the local magnetic field strength toward AGAL031 in Sect. \ref{sect:dis}. Finally, the main conclusions are summarized in Sect. \ref{sec:summary}.

\begin{table*}[b]
\caption{\label{tab:table1}Summary of the observations.}
\centering
\setlength{\tabcolsep}{3pt}
\begin{tabular}{ccccccccccc}
\hline
\hline
Date                  & Sources        & R.A.       & Decl.        & Cal. & Cal. & Cal. & N$_{ant}$ & D & $\theta_{\rm NH_3}$& $\theta_{\rm H_2O}$ \\
                      &                & (hh:mm:ss) & (dd:mm:ss)  & Flux       & Bandpass   & Phase      &           & (kpc)    & ($\arcsec$) &($\arcsec$ $\times$ $\arcsec$, $^{\circ}$)  \\
\hline
2021 Apr 06  & AGAL031.024+00.262 & 18:47:01.0 & -01:34:41.0 & 3C286 & J1743-0350 & J1832-1035 & 27 & 4.89 & 4.27 & 5.0$\times$2.8, -1.0\\
2021 Apr 08  & AGAL022.376+00.447 & 18:30:37.0 & -09:12:47.2 & 3C286 & J1743-0350 & J1832-1035 & 27 & 4.74  & 5.29&  6.7$\times$3.3, -39.3\\
2021 Mar 23 & AGAL024.314+00.086 & 18:35:19.0 & -07:37:27.2 & 3C286 & J1743-0350 & J1832-1035 & 25 & 6.54  & 5.89&  8.6$\times$5.2, 2.2\\
2021 Mar 30  & AGAL016.418-00.634 & 18:22:57.4 & -14:57:08:6 & 3C286 & J1743-0350 & J1832-1035 & 25 & \textcolor{black}{3.20} & 4.97 & 5.3$\times$3.0, 14.6\\
2021 Jun 01  & AGAL028.272-00.167 & 18:43:31.3 & -04:13:20.8 & 3C286 & J1743-0350 & J1832-1035 & 25 & \textcolor{black}{4.37}  & 4.35& 4.2$\times$3.1, 7.4\\
\hline
\end{tabular}
\raggedright
\tablecomments{Columns: (1) Observing date; (2) Source name; (3–4) Right ascension and declination; (5–7) Flux, bandpass, and phase calibrators; (8) Number of antennas; (9)\textcolor{black}{Distance from the Sun, calculated using the method of \citet{2019ApJ...885..131R}, based on maser parallax measurements and kinematic distances}; (10) Circular beam size of \amm(1,1); (11) Beam size of water data. Following the table order, the sources are aliased as AGAL031, AGAL022, AGAL024, AGAL016, and AGAL028.}
\end{table*}
\section{ Observations and data reduction}
\label{sec:obs}

VLA K-band observations in the D-array configuration were carried out in March, April, and June 2021 (Project code: 21A-128, PI: Shanghuo Li). The water line was observed at a rest frequency of 22.2350798~GHz. Ammonia inversion transitions, NH$_{3}$(J,K)=(1,1) at 23.6944955~GHz and NH$_{3}$(J,K)=(2,2) at 23.7226333~GHz, were also simultaneously observed towards the five clumps. The 8~MHz bandwidth consisted of 1024 channels with a channel width of 7.812~kHz, corresponding to a velocity resolution of 0.1~\kms\ at 23.7~GHz in dual-polarization mode.

Data calibration and imaging were performed using the CASA 6.5.4 software package \citep{2022PASP..134k4501C}. The calibrators used for these observations are identical for the five clumps and are listed in Table \ref{tab:table1}. To achieve a high signal-to-noise ratio (S/N), we used natural weighting to image the water line and smoothed the velocity resolution to 0.3~\kms. The noise level ($\sigma$) is 5.5, 5.8, 3.8, 4.9, and 5.2~\jyb\ for AGAL016, AGAL022, AGAL024, AGAL028, and AGAL031, respectively. The synthesized beam parameters are provided in Table~\ref{tab:table1}. For AGAL031, which exhibits the strongest water line peak intensity of $\approx 0.5\text{ Jy}$, we also imaged the Stokes V emission by setting the \texttt{tclean} parameter \texttt{stokes='V'}. Further details can be found in \citealt{2026arXiv260616355O} (hereafter Paper I).

To confirm the water detection, we compare VLA observational regions with the coverage region of the Radio Ammonia Mid-plane Survey (RAMPS) Pilot Survey \citep{2018ApJS..237...27H}. Four clumps, AGAL022, AGAL024, AGAL028, and AGAL031, are observed by the Green Bank Telescope in the RAMPS  project with the 22 GHz water line, which have  beam size of$\sim$ 34\arcsec.

Following the procedure of \citetalias{2026arXiv260616355O}, we identified ammonia cores using the velocity-integrated intensity of the satellite
hyperfine groups of the \amm(1,1) line.
For AGAL016, where the \amm(1,1) emission is significantly affected by missing flux owing to the lack of short-baseline coverage, we integrated the emission over two
velocity intervals, 30--38~\kms\ and 41--52~\kms.
For AGAL028, for which no 1.3~mm continuum observations are available in this work, we integrated the \amm(1,1) emission over 68.4--75.9~\kms. Notably, these two clumps were not included in \citetalias{2026arXiv260616355O}, whereas the other three sources were analyzed in the study of \citetalias{2026arXiv260616355O}.

%%%%%%%%
\begin{table*}[!t]
\centering
\setlength{\tabcolsep}{2pt}
\caption{\label{tab:table2} Properties of the regions associated with the detected water masers.}
% \scriptsize
% \renewcommand{\arraystretch}{1.2}

% \begin{tabular}{ccccccccccc}
\begin{tabular*}{\textwidth}{@{\extracolsep{\fill}}ccccccccccc@{}}
\hline
\hline
Source & Reg. & Num. & I & $\Delta V$ & V$_{\rm LSR}$ & V$_{\rm ref}$ & $\delta V$ & T$_k$ & $\Delta V_{{\rm NH}_3}$ & $N_{{\rm NH}_3}$ \\
name & & & (mJy) & (km s$^{-1}$) & (km s$^{-1}$) & (km s$^{-1}$) & (km s$^{-1}$) & (K) & (km s$^{-1}$) & ($\log_{10}$ cm$^{-2}$) \\
\hline
AGAL016 & \#1 & V$_1$ & 57 & 0.62(0.07) & 38.46(0.03) & 40.82(0.05) & 2.36 & 14.0(1.1) & 1.67(0.10) & 15.13(0.05) \\
        & \#2 & V$_1$ & 19 & 0.51(0.20) & 23.12(0.07) & 40.82(0.05)$^{*}$ & 17.68 & \ldots & - & \ldots \\
AGAL022 & \#1 & V$_1$ & 13 & 0.55(0.22) & 92.58(0.08) & 85.11(0.08) & -7.47 & \ldots & 0.89(0.18) & \ldots \\
        & \#2 & V$_1$ & 15 & 0.83(0.35) & 89.84(0.14) & 85.42(0.09) & -4.42 & \ldots & 1.54(0.20) & \ldots \\
        & \#3 & V$_1$ & 14 & 0.78(0.28) & 40.38(0.11) & 85.26(0.15) & 44.88 & \ldots & 1.76(0.42) & \ldots \\
AGAL024 & \#1 & V$_1$ & 117 & 0.80(0.04) & 115.52(0.02) & 114.69(0.03) & -0.83 & 16.13(0.57) & 2.35(0.06) & 15.38(0.02) \\
AGAL028 & \#1 & V$_1$ & 99 & 3.54(0.39) & 64.60(0.16) & 80.60(0.07) & 16.00 & 13.60(1.50) & 1.81(0.14) & 14.96(0.08) \\
AGAL031 & \#1 & V$_1$ & 32 & 0.80(0.15) & 91.13(0.06) & 96.11(0.01) & 4.99 & 13.97(0.35) & 0.99(0.02) & 15.11(0.02) \\
        &      & V$_2$ & 499 & 0.80(0.02) & 93.66(0.01) & - & 2.45 & - & - & - \\
        &      & V$_3$ & \ldots & 1.05(0.34) & 94.83(0.15) & - & 1.28 & - & - & - \\
        &      & V$_4$ & 144 & 0.67(0.03) & 98.35(0.01) & - & -2.24 & - & - & - \\
        & \#2 & V$_1$ & 134 & 0.59(0.02) & 85.40(0.01) & 96.37(0.01) & 10.97 & 13.98(0.41) & 0.81(0.02) & 15.13(0.02) \\
        & \#3 & V$_1$ & 132 & 0.78(0.03) & 128.94(0.01) & 78.20(0.03) & -50.73 & \ldots & \textcolor{black}{0.26(0.08)} & \ldots \\
\hline
\end{tabular*}

\raggedright
\tablecomments{Columns 1--3 list the source name, region label (hereafter Reg.~\#1, Reg.~\#2, and Reg.~\#3), and velocity component (hereafter V$_1$, V$_2$, etc.). Columns 4--6 give the water maser peak intensity, deconvolved the full width at half maximum (FWHM), and central velocity, respectively. Column 8 gives the velocity offset relative to the velocity of \amm(1,1) main line, whose parameters are listed in Columns 7 and 9--11. The FWHMs in Columns 5 and 10 were deconvolved from the spectral channel width of 0.3~km~s$^{-1}$. The FWHM of \amm(1,1) main line toward AGAL031 Reg.~\#3 is not deconvolved. Quantities marked with an asterisk were adopted from Reg.~\#1 of AGAL016. \textcolor{black}{For each region, the parameters in the last three columns are based on the first velocity component.}}

\end{table*}
%%%%%%%%

\section{Results}
\label{sec:result}

Using VLA K-band observations, we detect ten water maser spots toward dense cores within five $70~\mu\mathrm{m}$ infrared dark clumps. All of the detected emission remains spatially unresolved at the angular resolution of our observations (Fig.~\ref{fig:fig1}). 
\textcolor{black}{ Three clumps with maser spots,  namely Reg. \#1 of AGAL024, Reg. \#1 of AGAL028, and the V$_{4}$ component toward Reg. \#1 of AGAL031, have water maser counterparts at similar velocities that are spatially unresolved at the clump scale in the RAMPS survey \citep{2018ApJS..237...27H}. In addition, the water maser counterpart toward AGAL028 was listed in the Curated ATCA Census of High-Mass Clumps Legacy Survey, with a beam size of $27.6^{\prime\prime} \times 1.7^{\prime\prime}$ \citep{2024PASA...41...79A}.}
%Three spots, \textcolor{black}{Reg. \#1 of AGAL024, Reg. \#1 of AGAL028 and Reg. \#1 of AGAL031 with the V$_4$ component,  have unresolved clump counterparts} included in the RAMPS survey \citep{2018ApJS..237...27H}, and the spot toward AGAL028 is also listed in the Curated ATCA Census of High-Mass Clumps Legacy Survey \citep{2024PASA...41...79A}. 
To the best of our knowledge, the remaining seven spots have not previously been reported in dense cores within five 70 $\mu$m infrared dark clumps.

The 22~GHz water line transition is generally understood to be collisionally pumped in dense gas \citep[e.g.,][]{2002A&A...390..289B,2011A&A...527A..41S}. For each spot, we extracted a spectrum from a circular aperture centered on the emission peak, with a diameter equal to the major axis of the synthesized beam in the water line data. We fitted each spectral profile with one or more Gaussian components using the \texttt{curve\_fit} routine in \texttt{scipy.optimize} \citep{2020NaMet..17..261V}. The derived physical parameters are listed in Table \ref{tab:table2}.

\textcolor{black}{The ten water maser spots exhibit thirteen velocity components, with peak flux densities ranging from 13 to 499 mJy and FWHMs ranging from 0.51 to 3.54 \kms\ after deconvolution of the spectral resolution of 0.3~\kms. The strongest water maser emission is detected toward Reg. \#1 in AGAL031, with a beam-averaged peak brightness temperature of 88 K. In nine of the ten spots, the velocity components have FWHMs below 1.0~\kms\ and are narrower than the corresponding \amm(1,1) main lines. The exception is the 3.54~\kms\ component toward Reg. \#1 in AGAL028, which is nearly twice as broad as the local \amm(1,1) line. The velocity offsets from the systemic velocities range from 0.83 to 50.73 \kms. The components toward Reg.\#3 in AGAL022 and Reg. \#3 in AGAL031 have offsets greater than 30~\kms, whereas all other components have offsets below 20~\kms. As the spots are unresolved, their intrinsic brightness temperatures may be substantially higher than the measured beam-averaged values. The predominantly narrow profiles and the two components with large velocity offsets are consistent with compact, collisionally pumped water emission associated with shocked gas, as discussed further in Sect.~\ref{sect:dis1}.}

\begin{table*}
\centering
\setlength{\tabcolsep}{2pt}
\caption{\label{tab:table3} Locations of the detected water masers relative to the GBT data, dense cores, and CO~2--1 outflows.}
% \scriptsize
% \renewcommand{\arraystretch}{1.2}
% \begin{tabular*}{\textwidth}{@{\extracolsep{\fill}}cccccccccc@{}}
\begin{tabular*}{\textwidth}{@{\extracolsep{\fill}}cccccccccc@{}}
\hline
\hline
Source  & Reg. &R.A.&Decl.& GBT data& Ammonia& Cont. & CO & $M_{\rm core}$ & $n$ 
\\
name    &      & (hh:mm:ss)   &(dd:mm:ss)  &water detection&core  &   &    & ($M_\odot$) &
($10^{5}\,{\rm cm}^{-3}$) \\
\hline
%8.1 
AGAL016 & \#1 & 18:22:58.1& -14:57:02.1& &\checkmark & \checkmark & \checkmark & \textcolor{black}{9.26} & \textcolor{black}{1.14} \\
        & \#2 & 18:22:59.3 & -14:56:59.7& & &       & \checkmark &       &      \\

AGAL022 & \#1 &18:30:37.6&-09:13:34.9&   &   &      &            &       &      \\
        & \#2 &18:30:38.1&-09:13:23.8&   &    &    &            &       &      \\
        & \#3 &18:30:38.7&-09:13:22.3&     &   &    &            &       &      \\

AGAL024 & \#1 &18:35:19.2&-07:37:30.5& \checkmark& \checkmark&\checkmark &            & 10.89 & 130.5 \\

AGAL028 & \#1 &18:43:29.2&-04:12:35.3& \checkmark           &            &       &       \\

AGAL031 & \#1 &18:47:01.5&-01:34:47.3&\checkmark & &\checkmark & \checkmark & 25.4  & 6.81 \\
        & \#2 &18:47:01.6&-01:34:39.0&  &  &\checkmark           &             & 5.17  & 1.64 \\
        & \#3 &18:46:59.3&-01:33:08.7& & &          &            &       &      \\

\hline
\end{tabular*}

\raggedright
\tablecomments{
Columns 1 and 2 list the source name and region label, respectively.
Columns 3 and 4 indicate \textcolor{black}{peak intensity location of water maser spots}, respectively. Columns 5--9 point out the counterparts of GBT observation, ammonia core, 1.3 mm continuum core, and CO 2-1 outflows. Columns 8 and 9 give
the gas mass and volume density of the associated 1.3~mm continuum
core. These values are derived from \citet{2019ApJ...886..130L}
using the updated kinetic temperatures of Table \ref{tab:table2}.
}

\end{table*}

To characterize the spatial relation between the water masers and dense gas structures, Fig.~\ref{fig:fig1} also shows the 1.3~mm continuum emission and the \amm(1,1) main line emission, with red stars marking the continuum intensity peaks.
The field of view of the SMA 1.3~mm continuum observations covers only about one-sixth of the area mapped by the VLA ammonia observations.
We identified ammonia cores using the inner satellite hyperfine components of the \amm(1,1) line, following the method of \citetalias{2026arXiv260616355O}. In total, 74 ammonia cores were identified across the five clumps, but only three of them are spatially associated with water maser spots. We also compared the positions of the ten water maser spots with the 1.3~mm continuum cores identified by \citet{2019ApJ...886..130L}. Among the 21 continuum cores, only four are associated with water maser spots. This comparison excludes AGAL028, for which 1.3~mm continuum observations are unavailable. All comparisons are present in Table \ref{tab:table3}.

Four water maser spots are spatially associated with the 1.3~mm continuum.
In these regions, the peak intensities of the \amm(1,1) main line coincide spatially with the continuum emission.
Three of these four water maser spots are additionally associated with ammonia cores, and their peak intensities exceed those of water maser spots without dense core associations by more than a factor of three.
Although the ammonia cores do not fully spatially coincide with all ten water maser spots, \amm(1,1) emission is detected toward nine of them.
For Reg.~\#2 in AGAL016, where \amm(1,1) emission is not detected, we adopted the systemic velocity of \amm(1,1) in Reg.~\#1, which is consistent with the CO~2--1 line velocity
derived from the SMA observations.

\begin{figure*}[b]
    \centering

    \begin{subfigure}{0.99\linewidth}
        \centering
        \includegraphics[width=\linewidth]
            {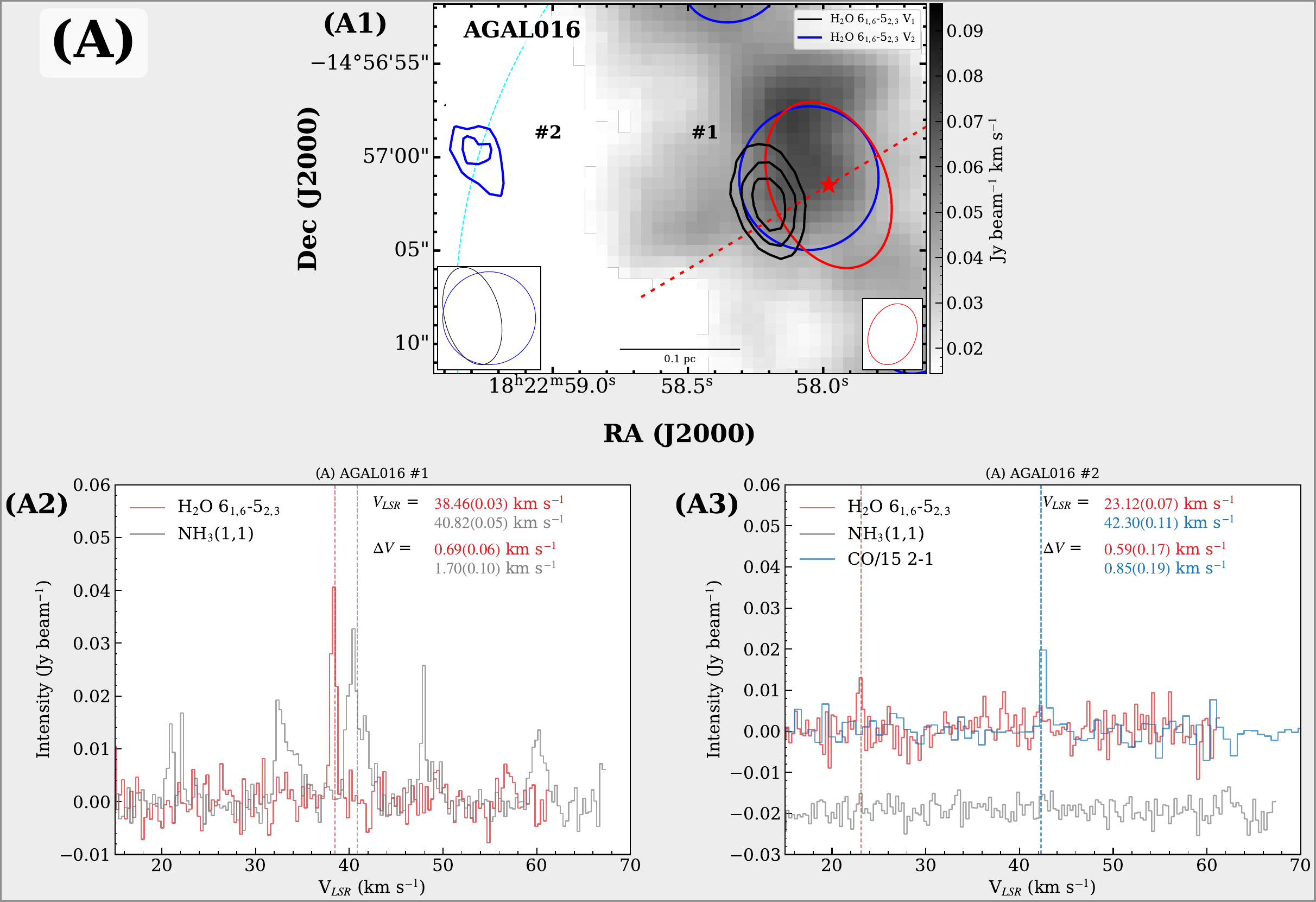}
        \caption{The adopted integration intervals are
    36.7--41.0~\kms\ for Reg.~\#1 and 22.6--23.8~\kms\ for Reg.~\#2.
    The blue and black contours start at 24.7 and
    9.0~m\jyb~\kms, with intervals of 12.3 and
    3.2~m\jyb~\kms, respectively.}
        \label{fig:fig1a}
    \end{subfigure}

    \caption{
    \textcolor{black}{Velocity integrated intensity maps of the \amm(1,1) main line
    for the five clumps are shown as gray scale backgrounds.
    The 1.3~mm continuum cores and ammonia cores are marked by red
    and blue ellipses, respectively. Red dashed lines indicate the
    CO 2--1 outflow axes reported by \citet{2019ApJ...886..130L}.Colored contours show the velocity integrated intensities of individual water maser components, while the annotations in each subfigure indicate the corresponding velocity integration ranges and contour intervals.
    The black circle represents the synthesized beam of the water line
    observations, while the red and blue circles indicate the beam sizes
    of the SMA 1.3~mm continuum and VLA \amm(1,1) observations,
    respectively. The gray and red spectra show the \amm(1,1) and water
    lines, respectively, and the dashed vertical lines mark the
    velocities. The colored annotations indicate the parameters derived from the line fitting, where the $\Delta V$ values not deconvolved from the channel width.}}

    \label{fig:fig1}
\end{figure*}

\begin{figure*}[!t]
    \ContinuedFloat
    \centering

    \begin{subfigure}{0.99\linewidth}
        \centering
        \includegraphics[width=\linewidth]
            {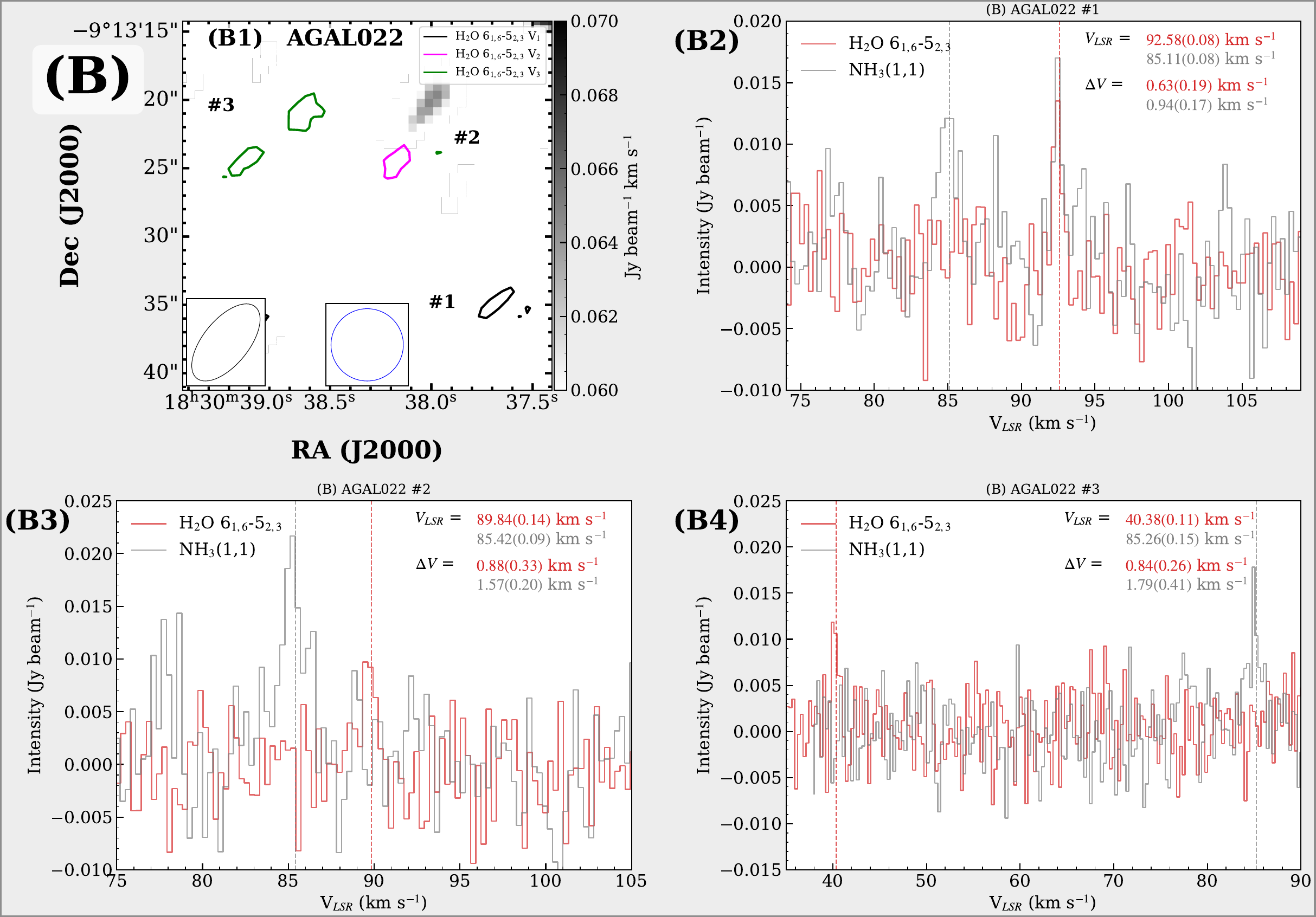}
        \caption{The intervals are 92.0--93.2~\kms\ for Reg.~\#1,
    89.0--90.5~\kms\ for Reg.~\#2, and 39.2--41.6~\kms\ for
    Reg.~\#3. The black, magenta, and green contours start at
    10.3, 11.6, and 14.6~m\jyb~\kms, respectively.}
        \label{fig:fig1b}
    \end{subfigure}

    \caption{Continued.}
\end{figure*}

\begin{figure*}[!t]
    \ContinuedFloat
    \centering

    \begin{subfigure}{0.99\linewidth}
        \centering
        \includegraphics[width=\linewidth]
            {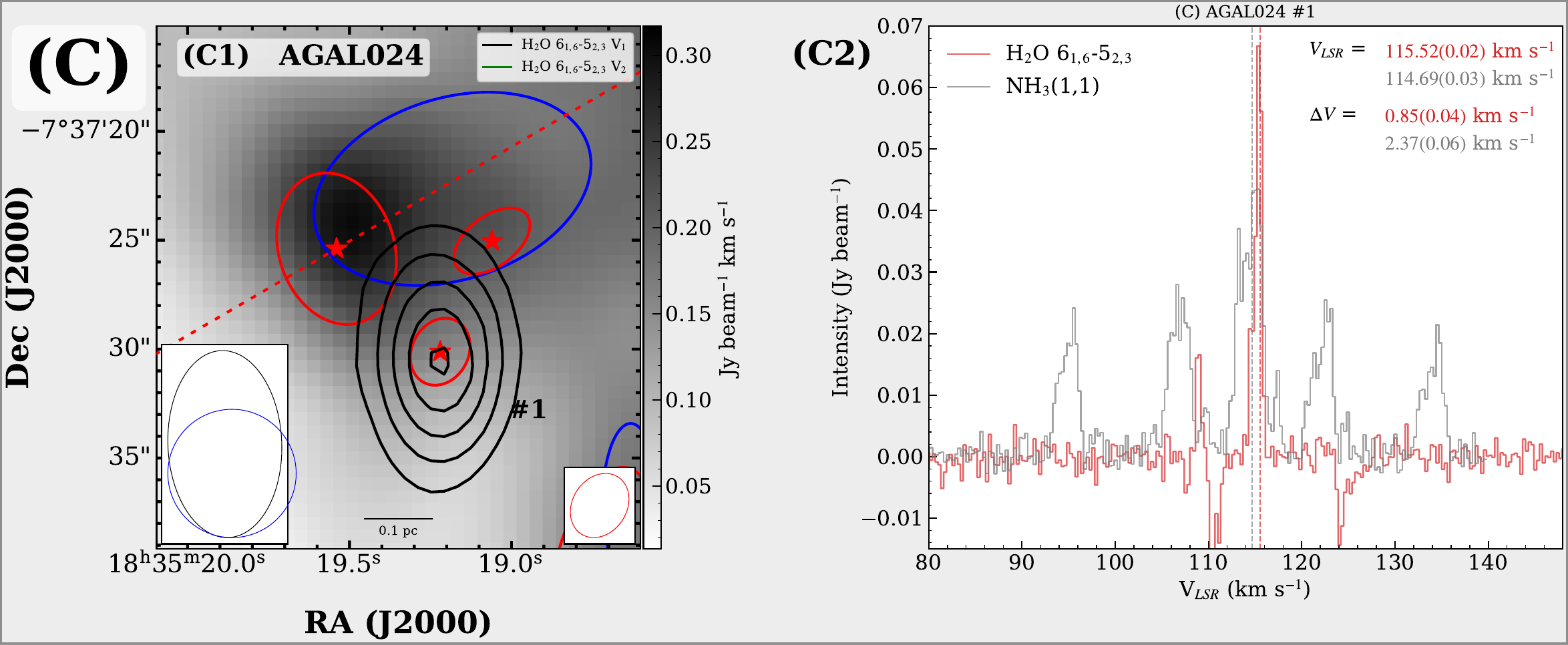}
        \caption{The interval is 113.0--117.0~\kms\ for Reg.~\#1.
    The black contours start at 28.7~m\jyb~\kms\ and increase in
    steps of 20.5~m\jyb~\kms.}
        \label{fig:fig1c}
    \end{subfigure}

    \vspace{1em}

    \begin{subfigure}{0.99\linewidth}
        \centering
        \includegraphics[width=\linewidth]
            {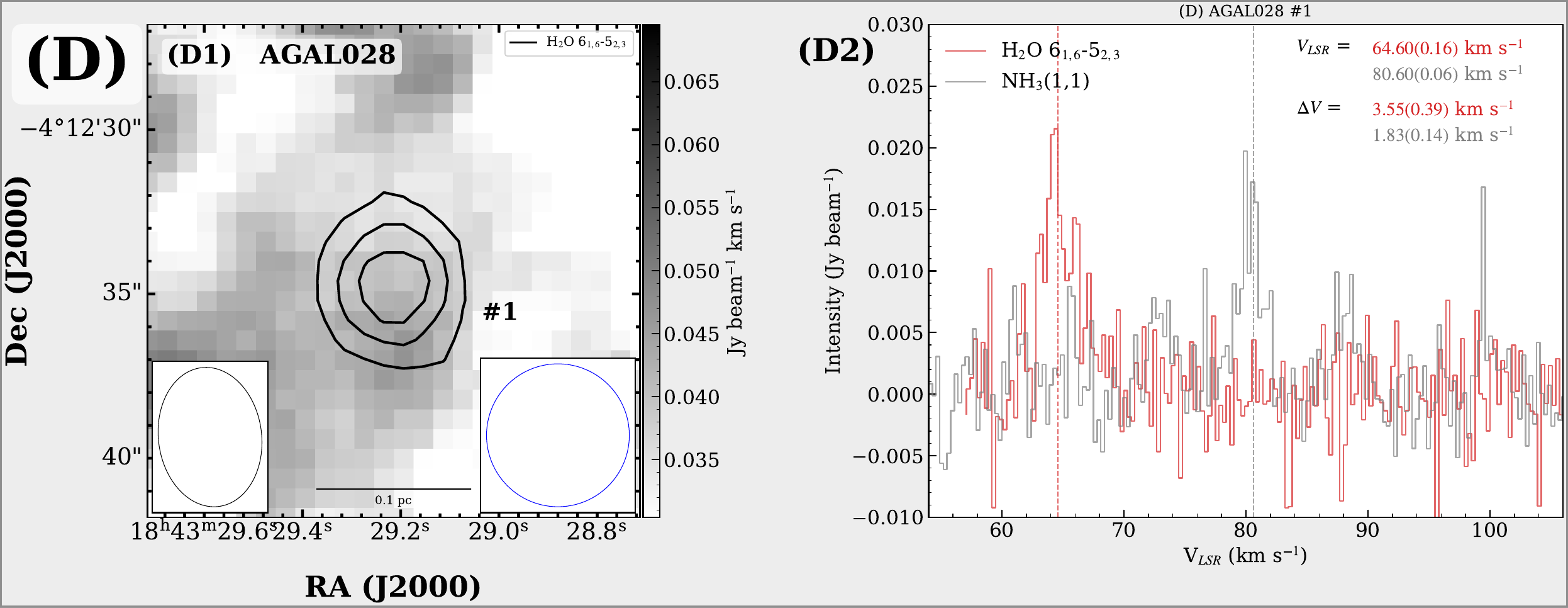}
        \caption{The interval is 60.0--69.0~\kms\ for Reg.~\#1.
    The black contours start at 31.9~m\jyb~\kms\ and increase in
    steps of 23.9~m\jyb~\kms.}
        \label{fig:fig1d}
    \end{subfigure}

    \caption{Continued.}
\end{figure*}

\begin{figure*}[t]
    \ContinuedFloat
    \centering

    \begin{subfigure}{0.99\linewidth}
        \centering
        \includegraphics[width=\linewidth]
            {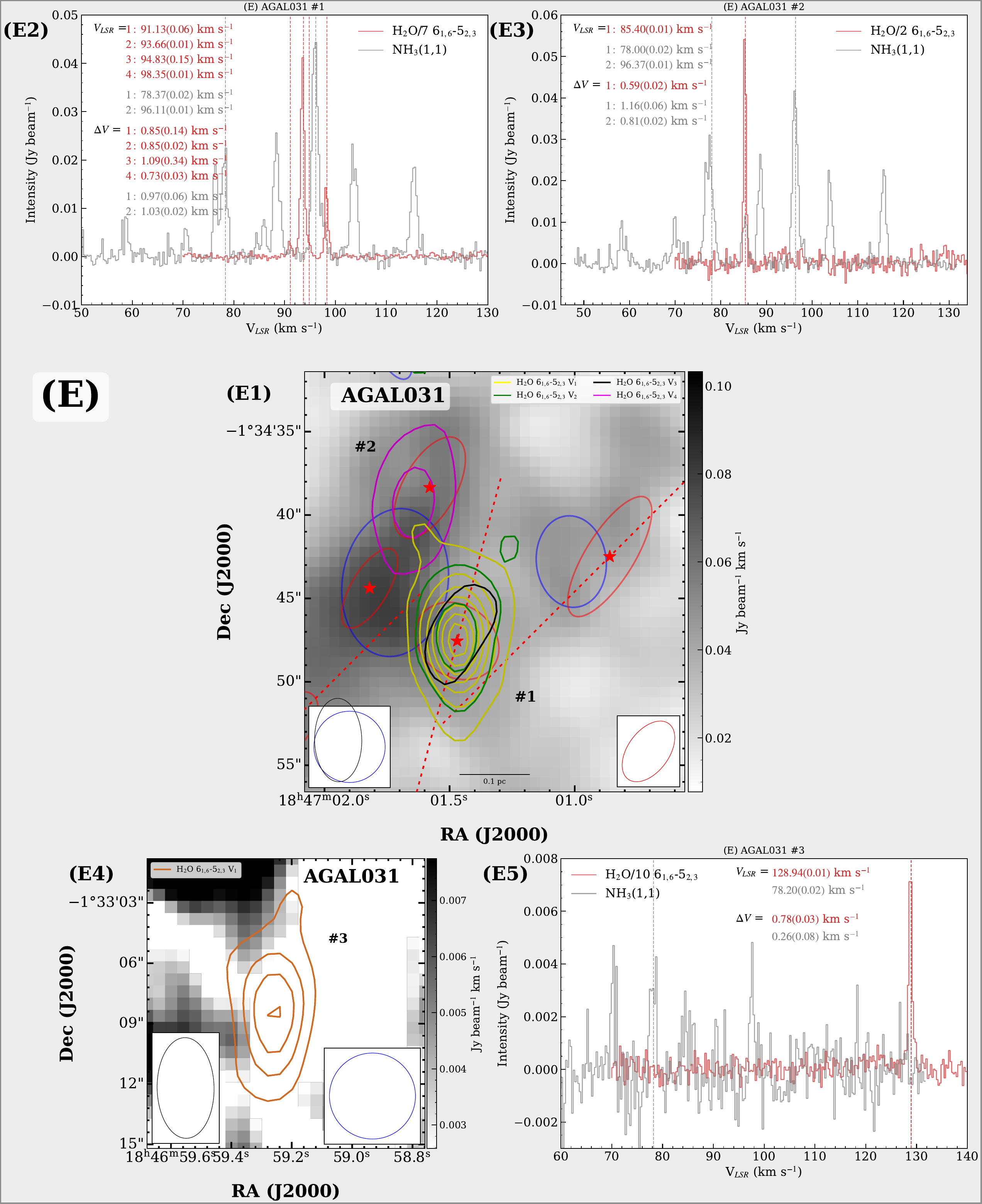}
        \caption{For Reg.~\#1, the adopted intervals are
    90.3--92.1, 92.1--96.3, and 96.3--100.2~\kms.
    Four velocity components are present, with the V$_2$ and V$_3$
    components integrated together. The intervals are
    83.7--87.0~\kms\ for Reg.~\#2 and 126.6--130.8~\kms\ for
    Reg.~\#3. The magenta, black, green, and yellow \water\ contours
    start at 20.6, 15.2, 22.4, and 23.3~m\jyb~\kms, with intervals
    of 62.2, 22.9, 67.4, and 69.9~m\jyb~\kms, respectively.
    The final panel shows the spectrum of Reg.~\#3 and the corresponding
    integrated-intensity contours, which start at
    23.3~m\jyb~\kms\ and increase in steps of
    35.0~m\jyb~\kms.}
        \label{fig:fig1e}
    \end{subfigure}
    \caption{Continued.}
\end{figure*}

\section{Discussion}
\label{sect:dis}
\subsection{The water maser spots associated with outflow}
\label{sect:dis1}

The water maser emission shows substantial variation, such as intensity and FWHM, among different velocity components, suggesting that the masers trace localized shocked gas rather than a single gas structure.
For example, the maser components in AGAL016 and AGAL022 show velocity offsets of up to 15~\kms\ and 52.2~\kms\ respect to the corresponding source  systemic velocity, respectively.
In AGAL016, the water maser spot in Reg. \#2 is located more than 0.1 pc away from both the ammonia core and the 1.3 mm continuum core, indicating that the maser emission is associated with gas impacted by local feedback or outflow-driven shocks.
Most of the detected maser spots show only one velocity component.
The exception is Reg.~\#1 of AGAL031, which exhibits four distinct components at 91.13~\kms, 93.66~\kms, 94.83~\kms, and 98.35~\kms\ (see panel e of Fig. \ref{fig:fig1}).
These components have similar FWHM of approximately 0.9~\kms\ and are distributed within a compact region.
Their intensities decrease from the outer part of the continuum core toward its center.
This compact multi-component structure may indicate that the maser emission is driven by the same embedded source through multiple ejection or shock events with different velocities.

Three continuum cores, Reg. \#1 in AGAL016 and Reg. \#1 and Reg. \#2 in AGAL031, exhibit outflows previously identified via CO 2-1 observations \citep{2019ApJ...886..130L}. The outflow direction is denoted by the red dashed line in Fig. \ref{fig:fig1}. Although there is a slight spatial offset between the peak intensities of the water maser and the 1.3~mm continuum emission, the water maser spots in Reg. \#1 of AGAL016 are located along the outflow axis. These outflows appear to traverse both the water masers and the dense cores. For the regions, the peak intensity locations of the water masers and the 1.3~mm continuum are well-aligned within a single synthesized beam, suggesting that the masers likely originate from shocks produced by the interaction of protostellar outflows with the surrounding dense gas.  We recalculated the masses and volume densities of the continuum cores using the kinetic temperatures listed in Column~9 of Table~\ref{tab:table2}, following the method of \citet{2019ApJ...886..130L}. These continuum cores have masses of \textcolor{black}{9.26}~$M_\odot$ and 25.4~$M_\odot$, with volume densities of \textcolor{black}{1.14}$\times$10$^{5}$~cm$^{-3}$ and 6.81$\times$10$^{5}$~cm$^{-3}$ for AGAL016 and AGAL031, respectively. \textcolor{black}{The typical uncertainty of gas mass and volume density is around 50\% according to \cite{2019ApJ...886..130L}}. This indicates that intermediate- to high-mass protostellar objects are most likely the driving sources of these shock-induced masers.

In contrast, no molecular outflow of CO 2-1 is detected toward the low-mass protostellar object in Reg. \#2 of AGAL031, which may indicate a weaker outflow or less advanced star formation activity. In Reg. \#1 of AGAL024, in addition to the 115.52 \kms\ component, two other potential maser features are detected near the edge of the VLA field of view. 
% These point sources near the edge of the field of view generate significant sidelobes in Reg. \#1 of AGAL024, resulting in artificial like features at 110~\kms\ and 124~\kms. To avoid spurious detections, we consider only the 115.52~\kms\ component to be a confirmed detection.
\textcolor{black}{These point sources near the edge of the field of view produce significant sidelobes in Reg. \#1 of AGAL024, resulting in spurious features at 110 \kms\ and 124 \kms. To avoid unreliable detections, we regard only the 115.52 \kms\ component as a confirmed detection.}
The confirmed water maser spot is associated with a continuum core that has a gas mass of 10.89~$M_\odot$. The volume density of this core reaches a maximum of 1.30$\times$10$^{7}$~cm$^{-3}$ with a peak maser intensity of 117~mJy, though no CO 2-1 outflow is detected at the current SMA spatial resolution of 3.4$\arcsec$. The water maser spot of AGAL028 is also not detected in the molecular outflow from CO 2-1 and SiO 5-4 reported by \cite{2017ApJ...841...97S}. This indicates that the water can serve as excellent probes for identifying  protostellar activity even in sources that remain quiescent in CO outflow emission. Thus, the water is highly complementary to tracers like CO in probing the protostellar environment, making them invaluable tools for investigating the extremely early stage of star formation.

\begin{figure}
    \centering
    \includegraphics[width=0.75\linewidth]{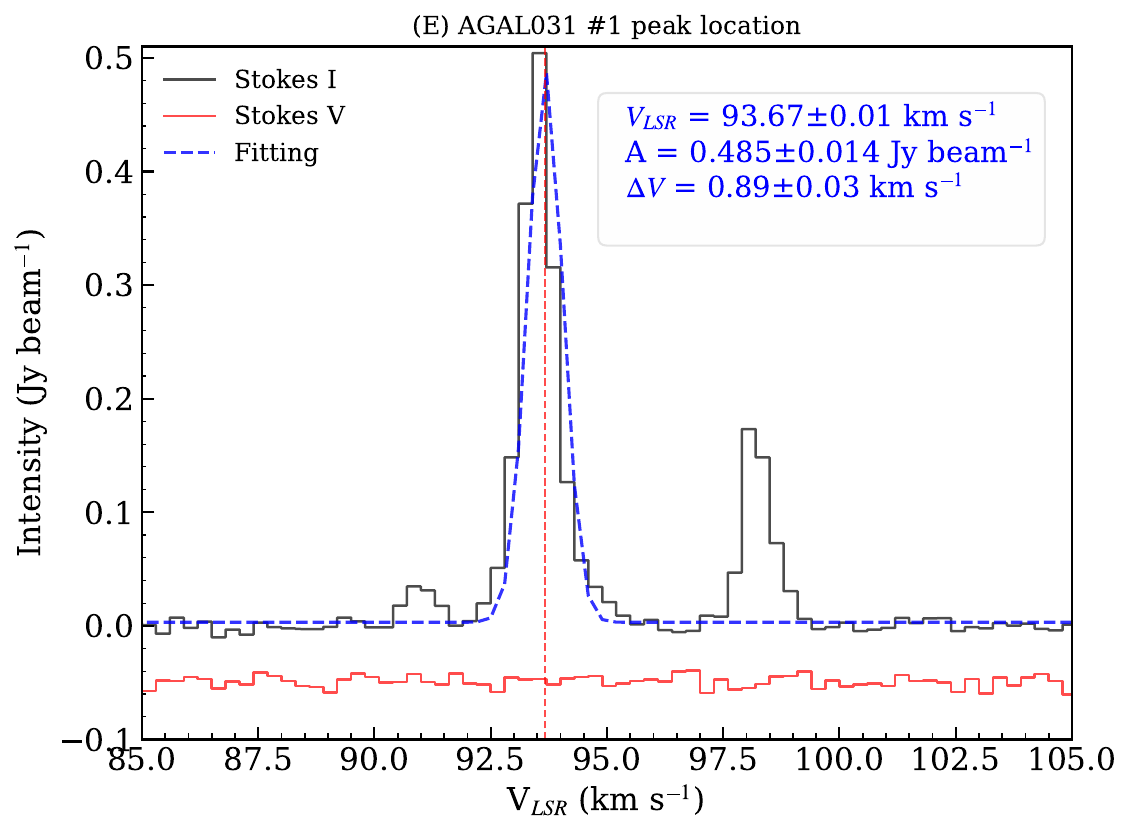}
    \caption{The figure presents the peak intensity spectra in Stokes I with a black line and Stokes V with a red line. The red dashed line denotes the velocity of the Stokes I line.}
    \label{fig:Fig2}
\end{figure}

\subsection{Water masers as a potential measuring tool of  magnetic fields}

The Zeeman effect remains the only direct observational method for measuring the line-of-sight magnetic field strength in molecular gas \citep{2019FrASS...6...66C}.
This method relies on the circular polarization signature encoded in the Stokes~I and Stokes~V spectra, where Stokes~I is defined as $(\mathrm{RCP}+\mathrm{LCP})/2$
and Stokes~V as $(\mathrm{RCP}-\mathrm{LCP})/2$.
Although no significant Stokes~V signal is detected in our VLA observations, the strongest water maser in Reg.~\#1 of AGAL031 provides a useful
constraint on the magnetic field strength.

\textcolor{black}{Following the methods of \citet{1989A&A...214..333F}
and \citet{2012A&A...542A..14A}, we used the brightest component of
the 22~GHz water line to estimate an upper limit on $B_{\mathrm{LOS}}$
using Equation~(2) of \cite{1989A&A...214..333F}:
\begin{equation}
\frac{T_{\mathrm{B}}(V)}{T_{\mathrm{B}}(I)}
= A_{F-F'}\,
  \frac{B_{\mathrm{LOS}}\,[\mathrm{G}]}
       {\Delta V\,[\mathrm{km\,s^{-1}}]},
\end{equation}
where $T_{\mathrm{B}}(V)$ and $T_{\mathrm{B}}(I)$ are the peak brightness
temperatures of the Stokes V and Stokes I spectra, respectively, and
$\Delta V$ is the FWHM of the Stokes I spectrum and the splitting
coefficient of $A_{F-F'}$.}
As shown in Fig.~\ref{fig:Fig2}, the V$_2$ component in the Stokes I
spectrum of Reg.~\#1 has a peak intensity of 489~mJy and an FWHM of
approximately 0.84~\kms. The rms noise levels are 3.3~m\jyb\ and
3.7~m\jyb\ for Stokes I and V, respectively. The peak Stokes V signal
has an S/N below 3, indicating that Zeeman splitting is not detected.
\textcolor{black}{Adopting the $1\sigma$ rms noise level of the Stokes V spectrum as an
upper limit on the Stokes V amplitude, and assuming $A_{F-F'}= 0.02$~\kms~G$^{-1}$
\citep[e.g.,][]{2006A&A...445.1031V,2012A&A...542A..14A}, we derive a upper limit of $B_{\mathrm{LOS}} \sim 318$~mG.
This upper limit provides only the magnitude of the line-of-sight
magnetic field component and  no information about its
orientation. Moreover, because it is derived from a single unresolved
maser component that likely traces shocked gas, it may not represent
the magnetic field strength of the host 70~$\mu$m infrared dark clump.}

Water maser emission is expected to arise in shocked gas with high volume densities, typically $n_{\rm H_2} \sim 10^8$--$10^{10}$~cm$^{-3}$
\citep{1989ApJ...346..983E}.
\textcolor{black}{At such densities, the value of $B_{\mathrm{LOS}}$ is larger than the magnetic field strengths inferred from water maser Zeeman observations in both low- and high-mass SFRs
\citep{2012A&A...542A..14A,2024ApJ...962..133R}.
Those studies detected much brighter Stokes~I emission, with peak intensities exceeding 100~Jy~beam$^{-1}$, and Stokes~V amplitudes of a few percent of Stokes~I.
By contrast, magnetic field strengths measured in prestellar cores using high critical density tracers, such as CCS~4--3, are typically only a few milligauss
\citep{2019A&A...631A..74M}.
Therefore, this upper limit does not provide a meaningful constraint on $B_{\mathrm{LOS}}$. Nevertheless, our study provides candidates for future measurements of magnetic field strengths in 70 $\mu$m infrared dark clumps.
Future high sensitivity and high angular resolution observations, particularly with the next-generation VLA and SKA, will be required to detect Stokes~V emission from the water maser spots identified in this work and to directly measure their magnetic field strengths.}

\section{Summary}
\label{sec:summary}
We used VLA K-band observations of the 22~GHz water line to investigate water maser emission toward five 70~$\mu$m infrared dark clumps at an early evolutionary stage. We detected ten water maser spots with thirteen distinct velocity components, eleven of which have narrow FWHMs of $<$1 \kms\ and velocity offsets below 20~\kms\ relative to the systemic velocities traced by the associated NH$_3$(1,1) main lines. Three maser spots are associated with molecular outflows traced by CO~2--1, supporting an origin in outflow-driven shocks, while three other regions show no corresponding CO outflow signature within the SMA field of view. This indicates that water masers can trace deeply embedded protostellar activity before CO outflows become detectable. For the strongest maser component in AGAL031, the non-detection of Stokes~$V$ emission yields an upper limit on the line-of-sight magnetic field strength on the order of \textcolor{black}{a few hundred mG}. Our results demonstrate that \textcolor{black}{water serve} as an unique tracer for constraining both kinematics and magnetic fields properties of protostellar systems at extremely early evolutionary stages.

\begin{acknowledgments}
    % We thank the referee for the comments and suggestions that improved this work.
    We sincerely thank  Prof. Zhiyu Zhang, Lingrui Lin, and Yichen Sun for their help with data reduction. We also acknowledge helpful with Fengyao Zhu, Siqi Zheng, Shuting Lin, Qizhou Zhang, Bo Zhang, Thushara Pillai, and Fei Li. This work is supported by National Key R$\&$D Program of China under grant 2023YFA1608204, the National Natural Science Foundation of China grant 12550003 and the Guangxi Talent Programme (Highland of Innovation Talents). S.L. acknowledges support from the National SKA Program of China with No. 2025SKA0140100, “Double First-Class” Funding with No. 14912217, and National Natural Science Foundation of China (NSFC) grant with No. 13004007. 
\end{acknowledgments}

% \facilities{SMA, VLA, GBT}
% \software{CASA \citep{2022PASP..134k4501C}, MIRIAD \citep{1995ASPC...77..433S}, astropy \citep{2013A&A...558A..33A}, astrodendro \citep{2008ApJ...679.1338R},  pyspeckit \citep{2011ascl.soft09001G,2022AJ....163..291G}}
\bibliographystyle{aasjournal}
\bibliography{ref}

\end{document}